\documentclass[
reprint,
superscriptaddress,
 amsmath,amssymb,
 aps,
 prl,
floatfix,
]{revtex4-2}
\usepackage{graphicx}
\usepackage{dcolumn}
\usepackage{bm}
\usepackage{natbib}
\usepackage{url}

\begin{document}

\preprint{APS/123-QED}
\title{Integrated Observation of Isotope-Dependent 
Turbulence, Zonal Flow, and Turbulence-Driven Transport}

\author{S. Ohshima}
\affiliation{Institute of Advanced Energy, Kyoto University, Gokasyo, Uji, Kyoto 611-0011, Japan}
\affiliation{Department of Physics and Astronomy, University of California, Irvine, Irvine, CA 92697, USA}

\author{H. Okada}
\affiliation{Institute of Advanced Energy, Kyoto University, Gokasyo, Uji, Kyoto 611-0011, Japan}

\author{S. Kobayashi}
\affiliation{Institute of Advanced Energy, Kyoto University, Gokasyo, Uji, Kyoto 611-0011, Japan}

\author{S. Kado}
\affiliation{Institute of Advanced Energy, Kyoto University, Gokasyo, Uji, Kyoto 611-0011, Japan}

\author{T. Minami}
\affiliation{Institute of Advanced Energy, Kyoto University, Gokasyo, Uji, Kyoto 611-0011, Japan}

\author{F. Kin}
\affiliation{Institute of Advanced Energy, Kyoto University, Gokasyo, Uji, Kyoto 611-0011, Japan}

\author{S. Inagaki}
\affiliation{Institute of Advanced Energy, Kyoto University, Gokasyo, Uji, Kyoto 611-0011, Japan}


\author{S. Konoshima}
\affiliation{Institute of Advanced Energy, Kyoto University, Gokasyo, Uji, Kyoto 611-0011, Japan}

\author{T. Mizuuchi}
\affiliation{Institute of Advanced Energy, Kyoto University, Gokasyo, Uji, Kyoto 611-0011, Japan}

\author{K. Nagasaki}
\affiliation{Institute of Advanced Energy, Kyoto University, Gokasyo, Uji, Kyoto 611-0011, Japan}




\date{\today}

\begin{abstract}
To address the long-standing unresolved issue of “isotope effect” in plasma transport, this study investigates the hydrogen/deuterium (H/D) isotope dependence of a nonlinear turbulence system. The analysis focuses on the zonal flow (ZF) activity, turbulence properties, their nonlinear interaction, and resulting turbulent transport in a torus plasma. ZF activity, observed in low-density electron cyclotron heating plasmas in Heliotron J, is enhanced with increasing D gas fraction from 10$\%$ to 80$\%$. While the turbulence scale size in the edge region ($\rho\sim0.8$) is larger in D plasmas, the reduction and decoupling of fluctuations, associated with an enhanced ZF, results in beneficial impacts on turbulent transport, driven by an enhanced nonlinear coupling between the ZF and turbulence in D plasmas. These differences in the turbulence nature lead to the significant reduction of turbulence-induced transport observed in the D plasma. These comprehensive observations suggest that the isotope dependence on the turbulence system is essential for explaining the isotope effect on confinement improvement and is vital in predicting the performance of future fusion reactors.
\end{abstract}

\keywords{Zonal Flow, Isotope Effect, Turbulence Transport, Nonlinear Coupling, Heliotron J}
\maketitle



For decades, the favorable H/D isotope effect on plasma confinement, where Deuterium (D) plasmas exhibit better confinement than Hydrogen (H), has been a critical, long-standing unresolved issue in fusion plasma science \cite{hawryluk1998fusion, bessenrodt1993isotope,scott1995isotopic,tokar2004nature, Cordey_1999JET, Schissel_1989DIII-D,maggi2017isotope, garcia2018isotope, urano2012energy, yamada2019isotope}. The experimental results fundamentally contradict predictions from simple transport models, including Gyro-Bohm scaling \cite{tokar2004nature, yamada2019isotope}. The turbulent diffusion coefficient, $D \sim L_c^2/\tau_c$, combined with the observed scaling of the turbulent scale length ($L_c$) proportional to the ion gyro-radius ($\rho_s$), predicts a resultant degradation of turbulent transport in D plasma due to larger turbulent structures \cite{ramisch2005rho, mckee2001non, ottaviani1999gyro}. 

It has been hypothesized that the isotope effect can be attributed to a turbulence system involving zonal flow(ZF) activity \cite{watanabe2011effects, hahm2013isotopic, xu2013isotope, nakata2017isotope, garcia2018isotope, shen2016isotope,braun2009effect,pusztai2013turbulent, Qi_PRR_2024,bustos2015microturbulence} alongside its established roles in steady-state transport \cite{diamond2005zonal, fujisawa2008review, hillesheim2016stationary, fujisawa2004identification} and transient phenomena in fusion plasmas\cite{manz2012zonal, schmitz2012role, itoh2013assessment, team2011mean, estrada2011spatiotemporal, garbet2001turbulence}. These works discuss the isotope dependence of residual ZF\cite{hahm2013isotopic}, trapped electron mode(TEM) turbulence\cite{nakata2017isotope}, electromagnetic effect\cite{garcia2018isotope}, non-adiabatic electrons\cite{belli2020PRL}, impurity ions\cite{shen2016isotope}, fast ions\cite{bonanomi2019isotope}, and radial electric field\cite{watanabe2011effects}. Experimentally, the dependence of ZF on the H/D ratio has been observed in tokamaks (TEXTOR, ISTTOK and FT-2) and helical devices (TJ-II and Heliotron J), in line with the aforementioned context \cite{xu2013isotope, liu2015isotope, liu2016multi, Gurchenko_ppcf2016isotope, ohshima2021ppcf}.

While the experimental results partially support the hypothesis, there is a lack of critical experimental evidence that turbulent transport associated with turbulence and ZF responds to the H/D ratio and of its underlying mechanism. To the best of our knowledge, this is the first comprehensive and meticulous study that provides experimental evidence of the isotope effect on turbulence, ZF, their nonlinear interaction, and transport.

The experiment was conducted in a medium-sized helical-axis Heliotron device, Heliotron J, whose design is based on the concept of quasi-isodynamic (omnigeneous) optimization\cite{wakatani2000study, obiki2001first}. The device has the major and averaged minor radii of 1.2 m and 0.17 m, respectively, with B = 1.25 T on the axis. The coil configuration, with a single helical coil with the poloidal/toroidal periods of $L/M = 1/4$, two types of toroidal coils, and three sets of vertical coils, provides flexibility of magnetic configuration, and the feature has been devoted to various configuration optimization studies

%
%
A ZF, a radially localized, symmetric flow on a flux surface, is observed in the ECH plasmas in Heliotron J\cite{ohshima2014observation, ohshima2021ppcf} using a toroidal long-range correlation technique \cite{fujisawa2008review}. The correlation is defined as $C(\tau, r_{probe1}) = \int \tilde{\phi}_{probe1}(t, r) \tilde{\phi}_{Ref}(t+\tau, r_{Ref}) dt$.  To isolate the ZF component, the frequency component below 4 kHz was extracted using a numerical low-pass filter before applying the analysis, since a significant coherence between the signals was observed toroidally separated probes in the frequency range of $< 4$ kHz in the spectral analysis. Note that there is no correlation with the density or magnetic fluctuations in the frequency range. Radial correlation profiles at $\tau$ = $-0.1$ ms, $0.0$ ms, and $+0.1$ ms are shown in Figure 1(a) and the correlation decay in the radial direction exhibits a radially localized zonal structure, and the maximum correlation at $\tau$ = 0 ms indicates a toroidally symmetric structure (m/n = 0/0) on the flux surface. This localized, symmetric, and temporally evolving correlation profile strongly indicates the presence of a ZF.

Because the correlation profile does not directly reflect the ZF structure, an effective ZF amplitude was evaluated with a technique using amplitudes and coherence\cite{ohshima2006observation}. Figure 1(b) shows the effective amplitude profile, exhibiting a narrower localization and steeper gradient of the ZF as compared to the correlation profile shown in Fig. 1 (a). The profile shown here reflects an oscillatory electric field at the steep gradient induced by the ZF.

The observed ZF interacts with turbulence and modulates the turbulence amplitude. Figure 1(b) exhibits the cross-correlation between the ZF and the fluctuation amplitude in the frequency range of 10--40 kHz, based on an amplitude correlation technique \cite{crossley1992experimental, duncan1993amplitude, liu2009characterizations,
yamada2008anatomy}. The correlation has a negative value at -50 mm where the ZF is localized, namely, turbulence amplitude is anti-correlated with the ZF. The radial extent of the turbulence suppression is notably consistent with that of the ZF, implying a causal relationship of ZF-turbulence interaction, where the ZF growth contributes to turbulence suppression and a resultant beneficial impact on turbulent transport through their interaction.

 \begin{figure}[t] 
 \includegraphics[width=\linewidth]
 {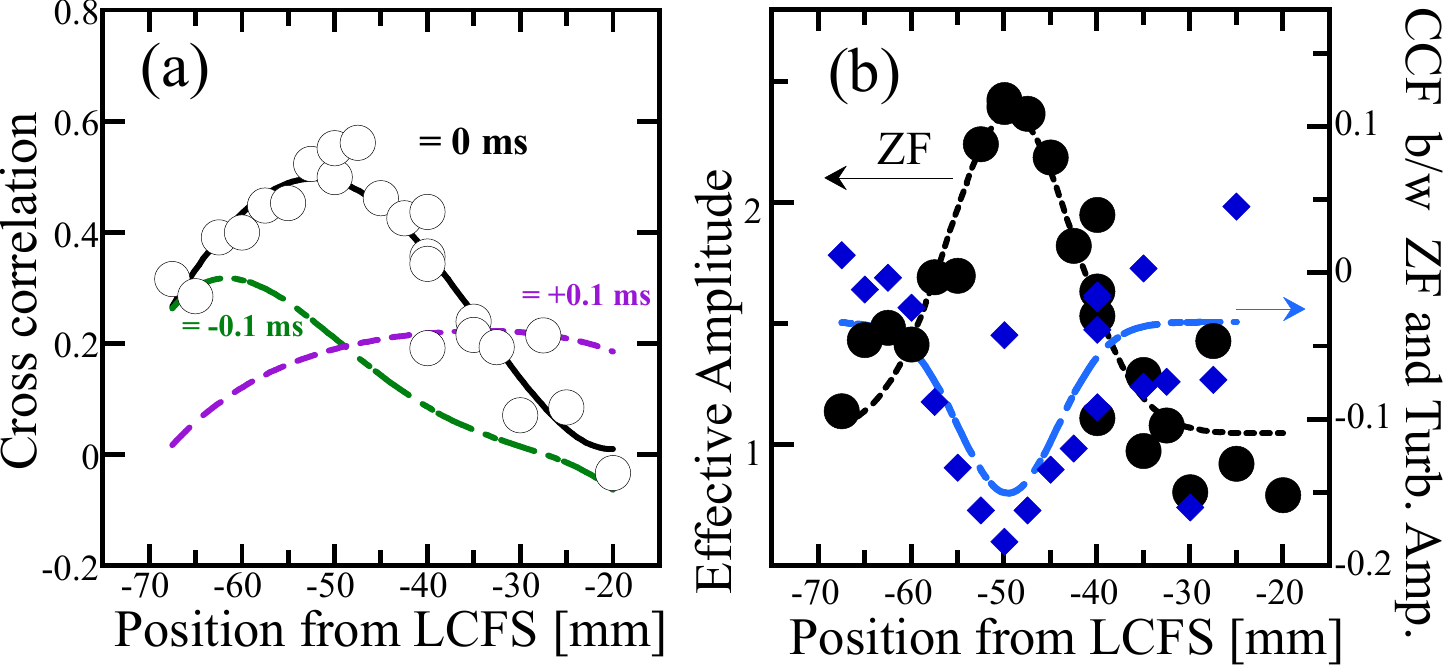} 
\caption{
Characterization of the Zonal Flow (ZF) and its coupling with turbulence. 
(a) Radial profiles of the cross-correlation at three time delays, $\tau = -0.1\,\mathrm{ms}$, $0\,\mathrm{ms}$, and $+0.1\,\mathrm{ms}$, obtained using the toroidal correlation technique between separated probes.
(b) Radial profiles of two key metrics: The effective ZF amplitude, evaluated from the fluctuation amplitude and coherence (black), and the anti-correlation between the ZF and the turbulence amplitude (blue).
} \label{Fig.1}
\end{figure}
%
%

The isotope mass ratio, defined as $= n_H/(n_D + n_H)$, was carefully controlled from $\sim0.1$ (D dominant) to $\sim 0.8$ (H dominant) based on the spectroscopic measurement, and the dependence of the ZF activity on the H/D ratio was examined in the experiment series, with two Langmuir probes fixed at $\rho \sim 0.8$.

A clear enhancement of the ZF activity from the H to D gas dominant discharges is observed, as shown in Fig. 2. The toroidal coherence and the ZF amplitude of $<$ 4 kHz are averaged and plotted as a function of the H/D ratio. The toroidal correlation and the ZF amplitude increase in proportion to the fraction of the D gas content, and vice versa. Both quantities systematically increase as the deuterium content rises: the averaged coherence increases from $\sim0.1-0.3$ to $\sim0.3-0.5$, and the effective ZF amplitude from $\sim 0.1-0.2$ to $\sim 0.4-0.5$, as the H/D ratio decreases from $\sim 0.8$ (H-dominant) to $\sim 0.1$ (D-dominant).  These results demonstrate that the low-frequency, toroidally symmetric flow is enhanced with increasing ion mass. This tendency is the same as the results in Tokamaks, while it contradicts the TJ-II stellarator result. Note that no confinement degradation was observed, and the stored energy remains at almost the same level even in the D plasmas of Heliotron J, which is inconsistent with the Gyro-Bohm scaling, as observed in the large helical device\cite{yamada2019isotope}. The observed ZF enhancement is suggested to constrain the confinement degradation caused by the increase of gyro-radius in the D plasmas.
\begin{figure}
\includegraphics[width=\linewidth]{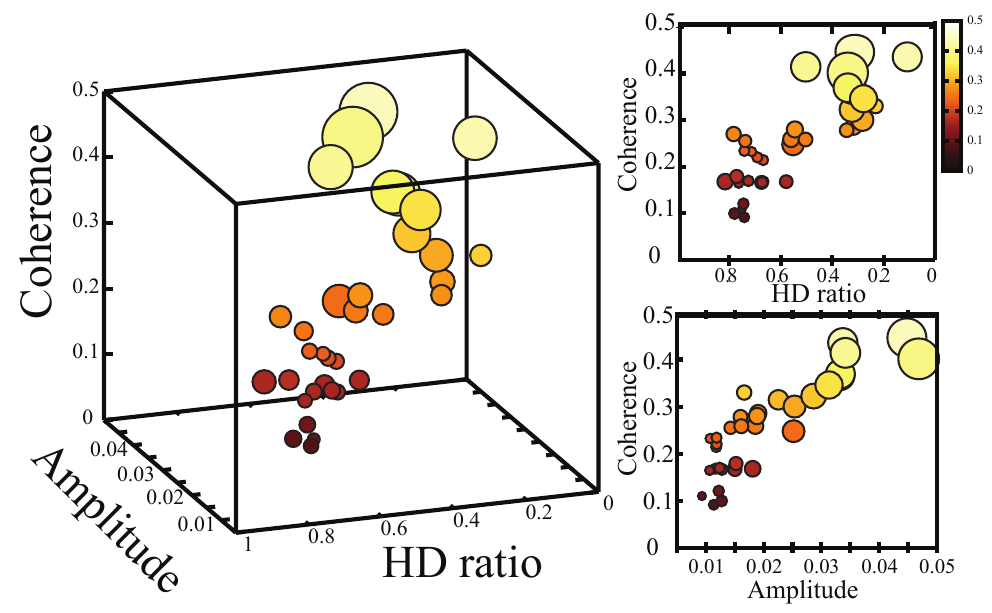}
\caption{
Isotope dependence of the ZF activity. Amplitude and coherence (long-range correlation) are plotted as a function of the H/D ratio, which was varied from $\sim 0.1$ (D-dominant) to $\sim 0.8$ (H-dominant). The ZF amplitude and coherence are also represented by the plot size and color, respectively. The inset figures on the right contain the same information as the three-dimensional figure but from different perspectives. Both the coherence and amplitude increase ($\sim$0.1–0.3 to $\sim$0.3–0.5 and $\sim$0.1–0.2 to $\sim$0.4–0.5, respectively) as the deuterium fraction increases.
}
\label{Fig.2}
\end{figure}
%

Turbulence responses against the variation of the isotope ratio and ZF activity are characterized in Fig. 3(a)--(h). The frequency spectra of the floating potential and ion saturation current fluctuations show that the turbulence level increases as H gas dominates, as shown in Fig. 3(a) and (b).  Higher frequency components emerge for potential fluctuation, whereas fluctuation levels increase in all frequency ranges for the ion saturation current in the H plasma.

As a factor to determine the turbulent transport level in a simple transport scaling, turbulence correlation length on the H/D ratio was investigated with a two-point cross-correlations technique. Probe tip pairs with a distance of approximately 5 mm were used to evaluate the correlation. Figures 3(c) and (d)  show the dependence of the two-point cross-correlation for the floating potential and the ion saturation current, respectively. Here, the dominant fluctuation in the frequency range from 10 to 40 kHz was extracted for the correlation analysis, which excludes a small crosstalk that could emerge in the high frequency range ($>$ 50 kHz) on the correlation analysis. 

While no clear dependence was seen for potential fluctuations, a significant isotope dependence was found for density fluctuations (ion saturation current). The peak value of cross-correlation for density fluctuation signals decreases as H gas becomes dominant. The value decreased from $\sim$ 0.35-0.45 in D to $\sim$ 0.1-0.2 in H, corresponding to the correlation length reduction from $\sim$ 0.4-0.6 mm in D to 0.2-0.3 mm in H, based on the assumption of exponential correlation decays as $ \propto exp(-\Delta/Lc)$. The ion mass scaling for turbulence size is consistent with past observations, and unfavorable for confinement based on the conventional transport framework \cite{ramisch2005rho, mckee2001non}.
\begin{figure}[t]
\includegraphics[width=\linewidth]
{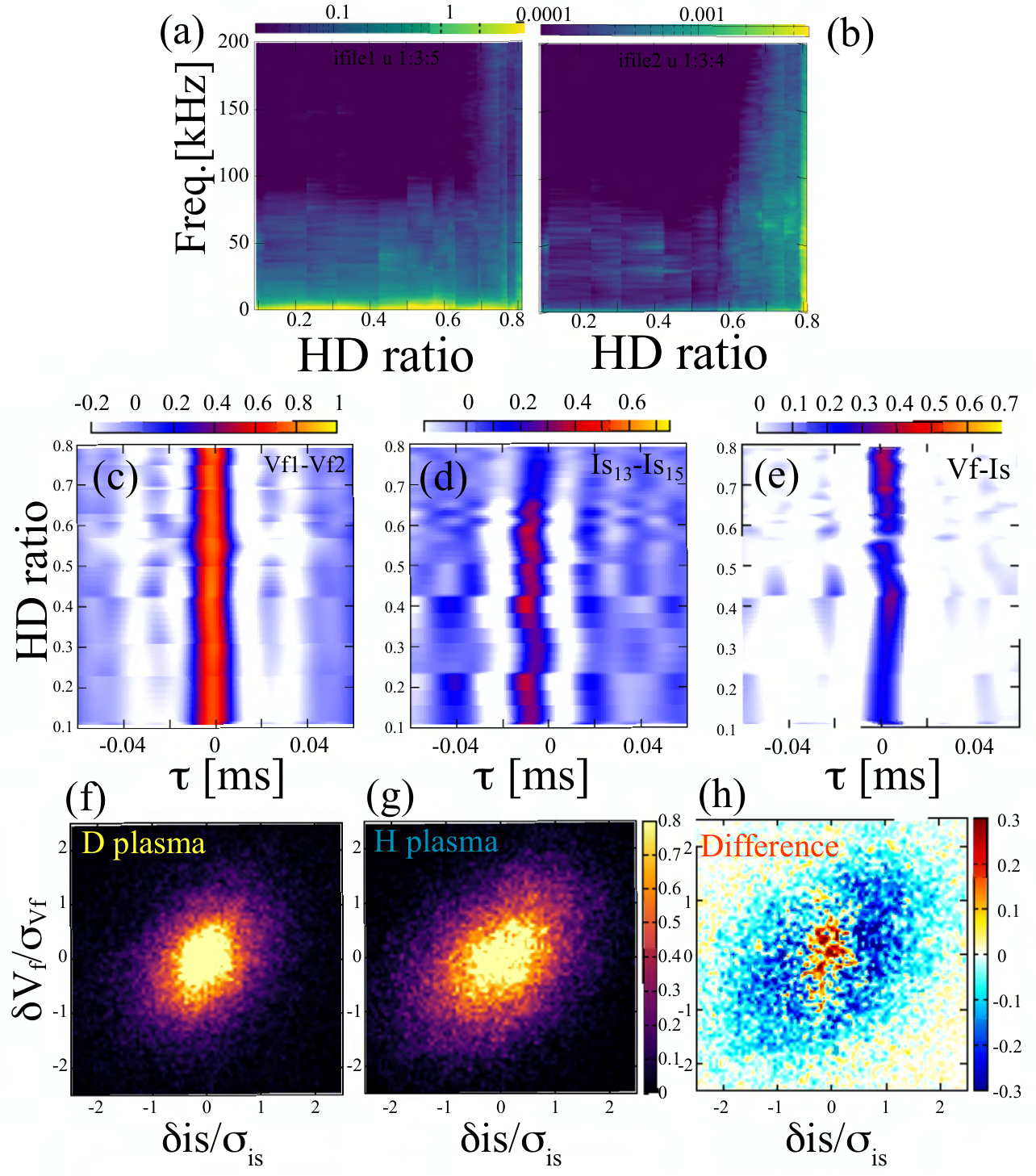}
\caption{
Isotope dependence of turbulence characteristics ($\mathrm{H}/\mathrm{D}$ ratio varied from $\sim 0.1$(D dominant) to $\sim 0.8$(H dominant). (a) Frequency spectra for potential fluctuation and (b) for density fluctuation.
(c) Two-point cross-correlation of floating potential and (d) ion saturation current (density fluctuation), measured between adjacent probes (5 mm separation). (e) Cross-correlation between potential and density fluctuations.
(f) Joint Probability Density Function in D plasma,  and (g) H plasma, $PDF_{D}$ and $PDF_{H}$, showing a round/elliptic shape indicating decoupled/correlated quantities(Is vs Vf).(h) Difference between the two PDF distributions ($PDF_{D} - PDF_{H}$), highlighting the reduction of the correlated, asymmetric component at large fluctuation amplitudes in D ( $>$ $\sigma$). Note that the axes are normalized by the standard deviation of fluctuation level to visualize the difference in the statistical properties of PDFs, and the H plasma has a broader distribution due to its larger fluctuation level without the normalization. 
}
\label{Fig.3}
\end{figure}
A correlation between different quantities is another factor to determine the turbulence-induced transport level, and hence, a cross-correlation between density and potential fluctuations was investigated, as shown in Fig. 3(e). The correlation decreases from ~0.4 to ~0.1 as the D gas is dominated. This decrement implies that the potential and density fluctuations are more decoupled in D plasmas. The observed decoupling contributes to the reduction of turbulence-induced transport, even if the turbulence scale size is larger in the D plasma.

The joint probability density function (PDF) provides a statistical measure of the coupling between two different quantities. To compare the statistical properties, the amplitudes were normalized by their respective standard deviations ($\sigma_{is}$, $\sigma_{Vf}$).
The joint-PDFs for the potential and density fluctuations have a more rounded distribution shape in D plasmas (Fig.3(f)), compared with the more elliptic shape in H plasmas (Fig.3(g)); The rounded shape indicates that the fluctuations are more decoupled between the fluctuations in D, whereas the elliptic shape suggests that they are correlated in H plasmas. The difference (Fig. 3(h)) between the two PDFs reveals that the correlated, asymmetric components are reduced at a higher fluctuation level ($> \sigma_{is}$ and $\sigma_{Vf}$) in D plasmas, while the uncorrelated components at a lower fluctuation level ( $< \sigma_{is}$, $\sigma_{Vf}$) increase. This reduction of the high-amplitude, correlated component is directly linked to the suppression of intermittent transport bursts in D plasma, demonstrating that the fluctuations exhibit a nature closer to random, uncorrelated noise. This favorable change in turbulence properties is consistent with the enhanced ZF activity in D plasmas.

\begin{figure}[t]
\includegraphics[width=\linewidth]
{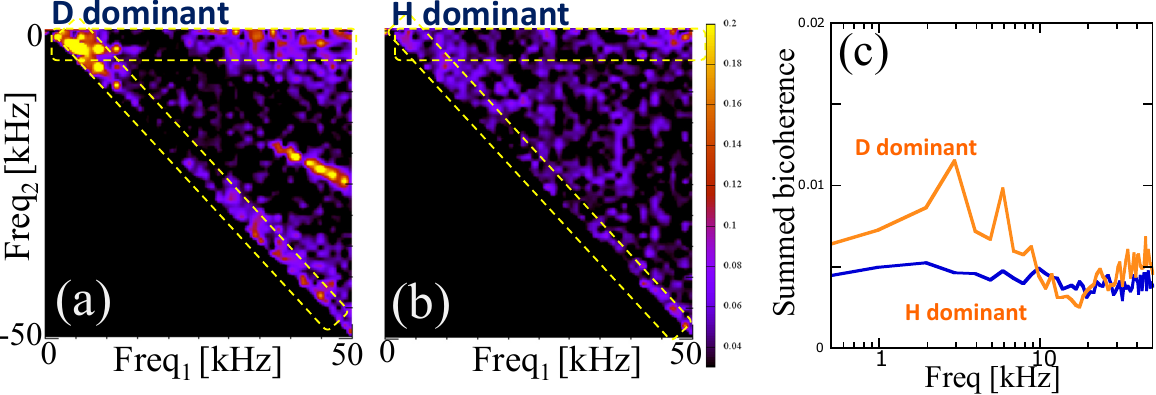}
\caption{Bicoherence analysis results for potential fluctuation in (a) D-dominant and (b) H-dominant discharges. (c) Spectra of summed bicoherence in the H and D dominant discharges.}
\label{Fig.4}
\end{figure}

An enhanced nonlinear interaction between the turbulence and ZF is indeed evident in D-dominant discharges. Figures 4(a) and (b) show the bicoherence results for the D and H dominant discharges, respectively. Auto-bicoherence is defined as $b^2(f_1,f_2) = |<X_{f_1} \cdot X_{f_2} \cdot X^{*}_{f_1+f_2} >|^2 /<|X_{f_1} \cdot X_{f_2}|^2> <|X_{f_1+f_2}|^2>$\cite{kim1979digital}\cite{van1995wavelet}\cite{tynan2001nonlinear}\cite{nagashima2005observation}. In the D-dominant case, nonlinear coupling in the low-frequency range of $\sim$ kHz and broadband turbulence is higher compared to that in the H dominant case. For simplicity of comparison, the summed bicoherence ($=\sum_{f=f_1 + f_2}b(f_1,f_2) $), which is an averaged bicoherence against each frequency component, is also shown in Fig. 4(c). 

In the D-dominant discharges, a clear increase in nonlinear coupling is observed in the frequency range below 4 kHz. The enhancement of this nonlinear coupling  in D plasmas leads to the larger energy transfer from the turbulence to the ZF mode and consequently results in the turbulent transport suppression through the enhanced ZF activity and the reduction and decoupling of the fluctuations.


The combined effects of the enhanced ZF activity, and turbulence suppression and decoupling result in the clear reduction of turbulence-induced transport in D plasmas, as demonstrated in Figure 5. Time-dependent particle flux is evaluated with ion saturation current and two floating potential signals, as $\Gamma \propto Is*(V_f1-V_f2)$ in Fig. 5(a). 

A one-dimensional PDF of $\Gamma$ characterizes that both distributions in H and D consistently have a distorted shape in positive $\Gamma$ value, namely, outward transport in Fig. 5(b). The substantially broader profile in H implies that stronger intermittent transport bursts. The average $\Gamma$ monotonically increases along with the H gas fraction, although with large scattering in Fig. 5(c). The $<\Gamma>$ increases by more than a factor of ~2 as the H fraction rose from 10 \% to 80 \%. Also, the amplitude of variation, $\delta\Gamma$, systematically depends on the H/D ratio in Fig. 5(d), exhibiting the enhanced turbulence transport activity in H plasmas. 

The observed significant reduction in $\Gamma$  and  $\delta \Gamma$ in D plasmas directly validates the mechanism of confinement improvement in D; the enhanced ZF(Fig.2)  and nonlinear coupling between ZF and turbulence(Fig.4) lead to the decoupling and suppression of high-amplitude, correlated turbulence(Fig. 3), thereby suppressing the intermittent turbulent transport(Fig.5).

\begin{figure}[t]
\includegraphics[width=\linewidth]
{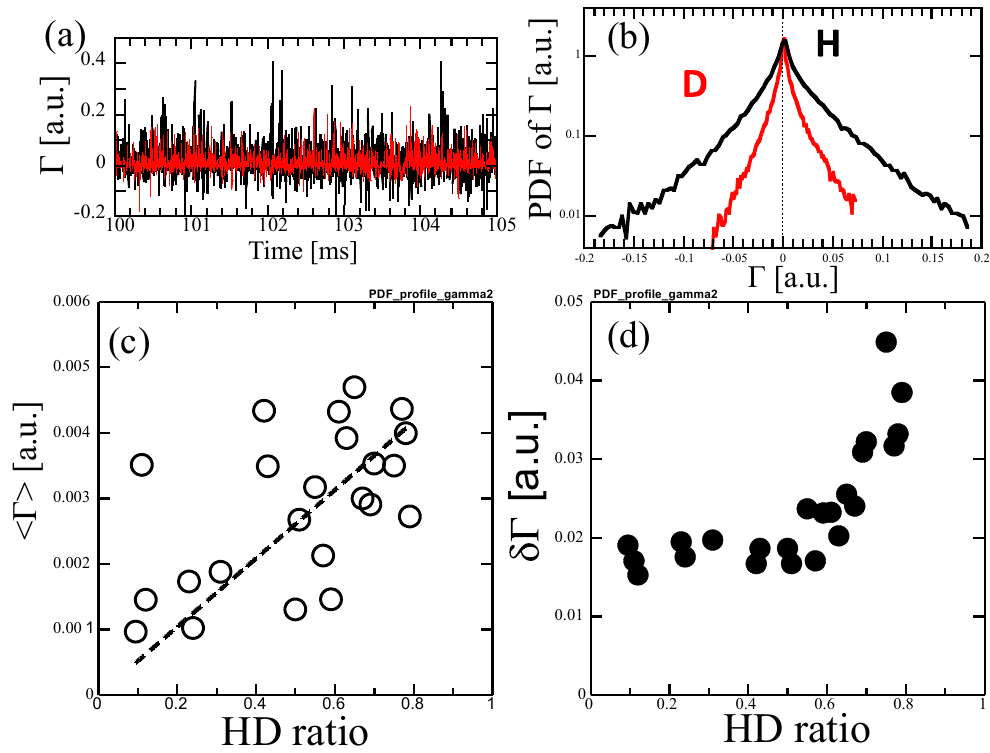}
\caption{Turbulence-induced transport properties versus H/D ratio varied from $\sim$ 0.1 (D-dominant) to $\sim$ 0.8 (H-dominant). (a)fluctuation-induced particle flux in D(red) and H(black), (b)one-dimensional PDF, isotope dependence of (c)average particle flux <$\Gamma$>, (d)fluctuation amplitude of particle flux $\delta\Gamma$ }
\label{Fig.5}
\end{figure}


In summary, the isotope dependence in a nonlinear turbulence system is characterized by low-density ECH plasmas in Heliotron J. Although the turbulence correlation length scales with ion mass, which predicts a worse confinement in D on the simple transport model, the enhanced ZF and its nonlinear coupling with turbulence introduce positive effects on the turbulence system, i.e., turbulence reduction and decoupling between fluctuations and subsequent turbulent transport suppression in D plasma. The observations are the first comprehensive evidence of the isotope effect on the nonlinear nature of the turbulence system and its transport in a torus plasma experiment. 

This analysis reveals the role of ZF and the nonlinearity of the turbulence system in the isotope effect on transport in a torus plasma. However, one should emphasize that other possible factors, such as impurities, neutrals, radial electric field, can be significant and should not be excluded under different plasma parameters, heating regimes, and operation schemes \cite{braun2009effect, watanabe2011effects, pusztai2013turbulent, shen2016isotope, belli2020PRL}, although they were likely not significant factors in our experiment\cite{ohshima2021ppcf}. In fact, confinement improvement factors are different among different machines, configurations, and operation schemes, suggesting that multiple factors can competitively and nonlinearly play roles on the isotope effect. For example, the result is in contrast to the recent theoretical work considering non-adiabatic electrons, which concludes that ZF has no critical role in the isotope effect. This discrepancy with this article suggests that the dominant mechanism of the isotope effect is regime-dependent, as has already been observed in many devices. The low-density ECH regime in this experiment allowed for the successful isolation of the isotope dependence of turbulence-ZF. 

This study strongly highlights the importance of multi-point measurement of fluctuations, correlation analyses among different quantities, statistical analysis, and higher-order analysis(e.g., bispectrum and bicoherence), which are expected to better facilitate exploring the isotope effect.

\subsection*{Acknowledgement}
The authors are grateful to the Heliotron J staff for their essential support during the experiments. One of the authors (SO) thanks Prof. Fujisawa for his continuous support, particularly for providing a part of the experimental instruments. This work was supported by the NIFS Collaboration Research program (NIFS10KUHL030, NIFS18KUHL084, NIFS18KUHL086, NIFS17KLPR039) and JSPS KAKENHI Grant Number 20K03901, 19H01875, 19K03802.

\bibliographystyle{unsrtnat}
\bibliography{test.bib}

\begin{thebibliography}{50}
\providecommand{\natexlab}[1]{#1}
\providecommand{\url}[1]{\texttt{#1}}
\expandafter\ifx\csname urlstyle\endcsname\relax
  \providecommand{\doi}[1]{doi: #1}\else
  \providecommand{\doi}{doi: \begingroup \urlstyle{rm}\Url}\fi

\bibitem[Hawryluk et~al.(1998)Hawryluk, Batha, Blanchard, Beer, Bell, Bell, Berk, Bernabei, Bitter, Breizman, et~al.]{hawryluk1998fusion}
RJ~Hawryluk, S~Batha, W~Blanchard, Michael Beer, MG~Bell, RE~Bell, H~Berk, S~Bernabei, M~Bitter, Boris Breizman, et~al.
\newblock Fusion plasma experiments on tftr: A 20 year retrospective.
\newblock \emph{Physics of Plasmas}, 5\penalty0 (5):\penalty0 1577--1589, 1998.

\bibitem[Bessenrodt-Weberpals et~al.(1993)Bessenrodt-Weberpals, Wagner, Gehre, Giannone, Hofmann, Kallenbach, McCormick, Mertens, Murmann, Ryter, et~al.]{bessenrodt1993isotope}
Monika Bessenrodt-Weberpals, Friedrich Wagner, O~Gehre, L~Giannone, JV~Hofmann, A~Kallenbach, K~McCormick, V~Mertens, HD~Murmann, F~Ryter, et~al.
\newblock The isotope effect in asdex.
\newblock \emph{Nuclear fusion}, 33\penalty0 (8):\penalty0 1205, 1993.

\bibitem[Scott et~al.(1995)Scott, Zarnstorff, Barnes, Bell, Bretz, Bush, Chang, Ernst, Fonck, Johnson, et~al.]{scott1995isotopic}
Steven~D Scott, Michael~C Zarnstorff, Cris~W Barnes, R~Bell, Norton~L Bretz, Charles Bush, Zuoyang Chang, D~Ernst, Raymond~J Fonck, L~Johnson, et~al.
\newblock Isotopic scaling of confinement in deuterium--tritium plasmas.
\newblock \emph{Physics of Plasmas}, 2\penalty0 (6):\penalty0 2299--2307, 1995.

\bibitem[Tokar et~al.(2004)Tokar, Kalupin, and Unterberg]{tokar2004nature}
MZ~Tokar, D~Kalupin, and B~Unterberg.
\newblock Nature of the isotope effect on transport in tokamaks.
\newblock \emph{Physical review letters}, 92\penalty0 (21):\penalty0 215001, 2004.

\bibitem[Cordey et~al.(1999)Cordey, Balet, Bartlett, Budny, Christiansen, Conway, Eriksson, Fishpool, Gowers, de~Haas, Harbour, Horton, Howman, Jacquinot, Kerner, Lowry, Monk, Nielsen, Righi, Rimini, Saibene, Sartori, Schunke, Sips, Smith, Stamp, Start, Thomsen, Tubbing, and von Hellermann]{Cordey_1999JET}
J.G Cordey, B~Balet, D.V Bartlett, R.V Budny, J.P Christiansen, G.D Conway, L.-G Eriksson, G.M Fishpool, C.W Gowers, J.C.M. de~Haas, P.J Harbour, L.D Horton, A.C Howman, J~Jacquinot, W~Kerner, C.G Lowry, R.D Monk, P~Nielsen, E~Righi, F.G Rimini, G~Saibene, R~Sartori, B~Schunke, A.C.C Sips, R.J Smith, M.F Stamp, D.F.H Start, K~Thomsen, B.J.D Tubbing, and M.G. von Hellermann.
\newblock Plasma confinement in {JET} h mode plasmas with h, d, {DT} and t isotopes.
\newblock \emph{Nuclear Fusion}, 39\penalty0 (3):\penalty0 301--308, mar 1999.
\newblock \doi{10.1088/0029-5515/39/3/301}.
\newblock URL \url{https://doi.org/10.1088%2F0029-5515%2F39%2F3%2F301}.

\bibitem[Schissel et~al.(1989)Schissel, Burrell, DeBoo, Groebner, Kellman, Ohyabu, Osborne, Shimada, Snider, Stambaugh, and and]{Schissel_1989DIII-D}
D.P. Schissel, K.H. Burrell, J.C. DeBoo, R.J. Groebner, A.G. Kellman, N.~Ohyabu, T.H. Osborne, M.~Shimada, R.T. Snider, R.D. Stambaugh, and T.S.~Taylor and.
\newblock Energy confinement properties of h-mode discharges in the {DIII}-d tokamak.
\newblock \emph{Nuclear Fusion}, 29\penalty0 (2):\penalty0 185--197, feb 1989.
\newblock \doi{10.1088/0029-5515/29/2/004}.
\newblock URL \url{https://doi.org/10.1088%2F0029-5515%2F29%2F2%2F004}.

\bibitem[Maggi et~al.(2017)Maggi, Weisen, Hillesheim, Chankin, Delabie, Horvath, Auriemma, Carvalho, Corrigan, Flanagan, et~al.]{maggi2017isotope}
CF~Maggi, H~Weisen, JC~Hillesheim, A~Chankin, E~Delabie, L~Horvath, F~Auriemma, IS~Carvalho, G~Corrigan, J~Flanagan, et~al.
\newblock Isotope effects on lh threshold and confinement in tokamak plasmas.
\newblock \emph{Plasma Physics and Controlled Fusion}, 60\penalty0 (1):\penalty0 014045, 2017.

\bibitem[Garcia et~al.(2018)Garcia, G{\"o}rler, and Jenko]{garcia2018isotope}
J~Garcia, T~G{\"o}rler, and F~Jenko.
\newblock Isotope and fast ions turbulence suppression effects: Consequences for high-$\beta$ iter plasmas.
\newblock \emph{Physics of Plasmas}, 25\penalty0 (5):\penalty0 055902, 2018.

\bibitem[Urano et~al.(2012)Urano, Takizuka, Fujita, Kamada, Nakano, Oyama, et~al.]{urano2012energy}
H~Urano, T~Takizuka, T~Fujita, Y~Kamada, T~Nakano, N~Oyama, et~al.
\newblock Energy confinement of hydrogen and deuterium h-mode plasmas in jt-60u.
\newblock \emph{Nuclear Fusion}, 52\penalty0 (11):\penalty0 114021, 2012.

\bibitem[Yamada et~al.(2019)Yamada, Tanaka, Seki, Suzuki, Ida, Fujii, Goto, Murakami, Osakabe, Tokuzawa, et~al.]{yamada2019isotope}
H~Yamada, K~Tanaka, R~Seki, C~Suzuki, K~Ida, K~Fujii, M~Goto, S~Murakami, M~Osakabe, T~Tokuzawa, et~al.
\newblock Isotope effect on energy confinement time and thermal transport in neutral-beam-heated stellarator-heliotron plasmas.
\newblock \emph{Physical review letters}, 123\penalty0 (18):\penalty0 185001, 2019.

\bibitem[Ramisch et~al.(2005)Ramisch, Mahdizadeh, Stroth, Greiner, Lechte, and Rahbarnia]{ramisch2005rho}
M~Ramisch, N~Mahdizadeh, U~Stroth, F~Greiner, C~Lechte, and K~Rahbarnia.
\newblock $\rho$ s scaling of characteristic turbulent structures in the torsatron tj-k.
\newblock \emph{Physics of plasmas}, 12\penalty0 (3):\penalty0 032504, 2005.

\bibitem[McKee et~al.(2001)McKee, Petty, Waltz, Fenzi, Fonck, Kinsey, Luce, Burrell, Baker, Doyle, et~al.]{mckee2001non}
GR~McKee, CC~Petty, RE~Waltz, C~Fenzi, RJ~Fonck, JE~Kinsey, TC~Luce, KH~Burrell, DR~Baker, EJ~Doyle, et~al.
\newblock Non-dimensional scaling of turbulence characteristics and turbulent diffusivity.
\newblock \emph{Nuclear Fusion}, 41\penalty0 (9):\penalty0 1235, 2001.

\bibitem[Ottaviani and Manfredi(1999)]{ottaviani1999gyro}
M~Ottaviani and G~Manfredi.
\newblock The gyro-radius scaling of ion thermal transport from global numerical simulations of ion temperature gradient driven turbulence.
\newblock \emph{Physics of Plasmas}, 6\penalty0 (8):\penalty0 3267--3275, 1999.

\bibitem[Watanabe et~al.(2011)Watanabe, Sugama, and Nunami]{watanabe2011effects}
TH~Watanabe, H~Sugama, and M~Nunami.
\newblock Effects of equilibrium-scale radial electric fields on zonal flows and turbulence in helical configurations.
\newblock \emph{Nuclear Fusion}, 51\penalty0 (12):\penalty0 123003, 2011.

\bibitem[Hahm et~al.(2013)Hahm, Wang, Wang, Yoon, and Duthoit]{hahm2013isotopic}
TS~Hahm, Lu~Wang, WX~Wang, ES~Yoon, and FX~Duthoit.
\newblock Isotopic dependence of residual zonal flows.
\newblock \emph{Nuclear Fusion}, 53\penalty0 (7):\penalty0 072002, 2013.

\bibitem[Xu et~al.(2013)Xu, Hidalgo, Shesterikov, Kr{\"a}mer-Flecken, Zoletnik, Van~Schoor, Vergote, Team, et~al.]{xu2013isotope}
Y~Xu, C~Hidalgo, I~Shesterikov, A~Kr{\"a}mer-Flecken, S~Zoletnik, M~Van~Schoor, M~Vergote, Textor Team, et~al.
\newblock Isotope effect and multiscale physics in fusion plasmas.
\newblock \emph{Physical review letters}, 110\penalty0 (26):\penalty0 265005, 2013.

\bibitem[Nakata et~al.(2017)Nakata, Nunami, Sugama, and Watanabe]{nakata2017isotope}
Motoki Nakata, Masanori Nunami, Hideo Sugama, and Tomo-Hiko Watanabe.
\newblock Isotope effects on trapped-electron-mode driven turbulence and zonal flows in helical and tokamak plasmas.
\newblock \emph{Physical review letters}, 118\penalty0 (16):\penalty0 165002, 2017.

\bibitem[Shen et~al.(2016)Shen, Dong, Sun, Qu, Lu, He, He, and Wang]{shen2016isotope}
Yong Shen, JQ~Dong, AP~Sun, HP~Qu, GM~Lu, ZX~He, HD~He, and LF~Wang.
\newblock Isotope effects of trapped electron modes in the presence of impurities in tokamak plasmas.
\newblock \emph{Plasma Physics and Controlled Fusion}, 58\penalty0 (4):\penalty0 045028, 2016.

\bibitem[Braun et~al.(2009)Braun, Helander, Belli, and Candy]{braun2009effect}
S~Braun, P~Helander, EA~Belli, and J~Candy.
\newblock Effect of impurities on collisional zonal flow damping in tokamaks.
\newblock \emph{Plasma Physics and Controlled Fusion}, 51\penalty0 (6):\penalty0 065011, 2009.

\bibitem[Pusztai et~al.(2013)Pusztai, Moll{\'e}n, F{\"u}l{\"o}p, and Candy]{pusztai2013turbulent}
Istv{\'a}n Pusztai, Albert Moll{\'e}n, T{\"u}nde F{\"u}l{\"o}p, and J~Candy.
\newblock Turbulent transport of impurities and their effect on energy confinement.
\newblock \emph{Plasma Physics and Controlled Fusion}, 55\penalty0 (7):\penalty0 074012, 2013.

\bibitem[Qi et~al.(2024)Qi, Kwon, Hahm, Leconte, Yi, Cho, and Seo]{Qi_PRR_2024}
Lei Qi, Jae-Min Kwon, T.~S. Hahm, M.~Leconte, Sumin Yi, Y.~W. Cho, and Janghoon Seo.
\newblock Role of isotopes in microturbulence from linear to saturated ohmic confinement regimes.
\newblock \emph{Phys. Rev. Res.}, 6:\penalty0 L012004, Jan 2024.
\newblock \doi{10.1103/PhysRevResearch.6.L012004}.
\newblock URL \url{https://link.aps.org/doi/10.1103/PhysRevResearch.6.L012004}.

\bibitem[Bustos et~al.(2015)Bustos, Ba{\~n}{\'o}n~Navarro, G{\"o}rler, Jenko, and Hidalgo]{bustos2015microturbulence}
A~Bustos, A~Ba{\~n}{\'o}n~Navarro, T~G{\"o}rler, F~Jenko, and C~Hidalgo.
\newblock Microturbulence study of the isotope effect.
\newblock \emph{Physics of Plasmas}, 22\penalty0 (1):\penalty0 012305, 2015.

\bibitem[Diamond et~al.(2005)Diamond, Itoh, Itoh, and Hahm]{diamond2005zonal}
Patrick~H Diamond, SI~Itoh, K~Itoh, and TS~Hahm.
\newblock Zonal flows in plasma—a review.
\newblock \emph{Plasma Physics and Controlled Fusion}, 47\penalty0 (5):\penalty0 R35, 2005.

\bibitem[Fujisawa(2008)]{fujisawa2008review}
Akihide Fujisawa.
\newblock A review of zonal flow experiments.
\newblock \emph{Nuclear Fusion}, 49\penalty0 (1):\penalty0 013001, 2008.

\bibitem[Hillesheim et~al.(2016)Hillesheim, Delabie, Meyer, Maggi, Meneses, Poli, Contributors, Consortium, et~al.]{hillesheim2016stationary}
JC~Hillesheim, E~Delabie, H~Meyer, CF~Maggi, L~Meneses, E~Poli, JET Contributors, EUROfusion Consortium, et~al.
\newblock Stationary zonal flows during the formation of the edge transport barrier in the jet tokamak.
\newblock \emph{Physical review letters}, 116\penalty0 (6):\penalty0 065002, 2016.

\bibitem[Fujisawa et~al.(2004)Fujisawa, Itoh, Iguchi, Matsuoka, Okamura, Shimizu, Minami, Yoshimura, Nagaoka, Takahashi, et~al.]{fujisawa2004identification}
A~Fujisawa, K~Itoh, H~Iguchi, K~Matsuoka, S~Okamura, A~Shimizu, T~Minami, Y~Yoshimura, K~Nagaoka, C~Takahashi, et~al.
\newblock Identification of zonal flows in a toroidal plasma.
\newblock \emph{Physical review letters}, 93\penalty0 (16):\penalty0 165002, 2004.

\bibitem[Manz et~al.(2012)Manz, Xu, Wan, Wang, Guo, Cziegler, Fedorczak, Holland, M{\"u}ller, Thakur, et~al.]{manz2012zonal}
P~Manz, GS~Xu, BN~Wan, HQ~Wang, HY~Guo, I~Cziegler, N~Fedorczak, C~Holland, SH~M{\"u}ller, SC~Thakur, et~al.
\newblock Zonal flow triggers the lh transition in the experimental advanced superconducting tokamak.
\newblock \emph{Physics of Plasmas}, 19\penalty0 (7):\penalty0 072311, 2012.

\bibitem[Schmitz et~al.(2012)Schmitz, Zeng, Rhodes, Hillesheim, Doyle, Groebner, Peebles, Burrell, and Wang]{schmitz2012role}
L~Schmitz, L~Zeng, TL~Rhodes, JC~Hillesheim, EJ~Doyle, RJ~Groebner, WA~Peebles, KH~Burrell, and G~Wang.
\newblock Role of zonal flow predator-prey oscillations in triggering the transition to h-mode confinement.
\newblock \emph{Physical review letters}, 108\penalty0 (15):\penalty0 155002, 2012.

\bibitem[Itoh et~al.(2013)Itoh, Itoh, and Fujisawa]{itoh2013assessment}
Kimitaka Itoh, Sanae-I Itoh, and Akihide Fujisawa.
\newblock An assessment of limit cycle oscillation dynamics prior to lh transition.
\newblock \emph{Plasma and Fusion Research}, 8:\penalty0 1102168--1102168, 2013.

\bibitem[Team et~al.(2011)Team, Conway, Angioni, Ryter, Sauter, and Vicente]{team2011mean}
ASDEX~Upgrade Team, GD~Conway, C~Angioni, F~Ryter, P~Sauter, and J~Vicente.
\newblock Mean and oscillating plasma flows and turbulence interactions across the l- h confinement transition.
\newblock \emph{Physical review letters}, 106\penalty0 (6):\penalty0 065001, 2011.

\bibitem[Estrada et~al.(2011)Estrada, Hidalgo, Happel, and Diamond]{estrada2011spatiotemporal}
T~Estrada, C~Hidalgo, T~Happel, and PH~Diamond.
\newblock Spatiotemporal structure of the interaction between turbulence and flows at the lh transition in a toroidal plasma.
\newblock \emph{Physical review letters}, 107\penalty0 (24):\penalty0 245004, 2011.

\bibitem[Garbet(2001)]{garbet2001turbulence}
X~Garbet.
\newblock Turbulence in fusion plasmas: key issues and impact on transport modelling.
\newblock \emph{Plasma physics and controlled fusion}, 43\penalty0 (12A):\penalty0 A251, 2001.

\bibitem[Belli et~al.(2020)Belli, Candy, and Waltz]{belli2020PRL}
E.~A. Belli, J.~Candy, and R.~E. Waltz.
\newblock Reversal of simple hydrogenic isotope scaling laws in tokamak edge turbulence.
\newblock \emph{Phys. Rev. Lett.}, 125:\penalty0 015001, Jun 2020.
\newblock \doi{10.1103/PhysRevLett.125.015001}.
\newblock URL \url{https://link.aps.org/doi/10.1103/PhysRevLett.125.015001}.

\bibitem[Bonanomi et~al.(2019)Bonanomi, Casiraghi, Mantica, Challis, Delabie, Fable, Gallart, Giroud, Lerche, Lomas, Menmuir, Staebler, Taylor, Van~Eester, and contributors]{bonanomi2019isotope}
N.~Bonanomi, I.~Casiraghi, P.~Mantica, C.~Challis, E.~Delabie, E.~Fable, D.~Gallart, C.~Giroud, E.~Lerche, P.~Lomas, S.~Menmuir, G.M. Staebler, D.~Taylor, D.~Van~Eester, and JET contributors.
\newblock Role of fast ion pressure in the isotope effect in jet l-mode plasmas.
\newblock \emph{Nuclear Fusion}, 59\penalty0 (9):\penalty0 096030, jul 2019.
\newblock \doi{10.1088/1741-4326/ab2d4f}.
\newblock URL \url{https://doi.org/10.1088/1741-4326/ab2d4f}.

\bibitem[Liu et~al.(2015)Liu, Pedrosa, van Milligen, Hidalgo, Silva, Tabar{\'e}s, Zurro, McCarthy, Cappa, Liniers, et~al.]{liu2015isotope}
B~Liu, MA~Pedrosa, B~Ph van Milligen, C~Hidalgo, C~Silva, FL~Tabar{\'e}s, B~Zurro, KJ~McCarthy, A~Cappa, M~Liniers, et~al.
\newblock Isotope effect physics, turbulence and long-range correlation studies in the tj-ii stellarator.
\newblock \emph{Nuclear Fusion}, 55\penalty0 (11):\penalty0 112002, 2015.

\bibitem[Liu et~al.(2016)Liu, Silva, Figueiredo, Pedrosa, van Milligen, Pereira, Losada, and Hidalgo]{liu2016multi}
B~Liu, C~Silva, H~Figueiredo, MA~Pedrosa, B~Ph van Milligen, T~Pereira, U~Losada, and C~Hidalgo.
\newblock Multi-scale study of the isotope effect in isttok.
\newblock \emph{Nuclear Fusion}, 56\penalty0 (5):\penalty0 056012, 2016.

\bibitem[Gurchenko et~al.(2016)Gurchenko, Gusakov, Niskala, Altukhov, Esipov, Kiviniemi, Korpilo, Kouprienko, Lashkul, Leerink, Perevalov, and Irzak]{Gurchenko_ppcf2016isotope}
A~D Gurchenko, E~Z Gusakov, P~Niskala, A~B Altukhov, L~A Esipov, T~P Kiviniemi, T~Korpilo, D~V Kouprienko, S~I Lashkul, S~Leerink, A~A Perevalov, and M~A Irzak.
\newblock The isotope effect in turbulent transport control by gams. observation and gyrokinetic modeling.
\newblock \emph{Plasma Physics and Controlled Fusion}, 58\penalty0 (4):\penalty0 044002, jan 2016.
\newblock \doi{10.1088/0741-3335/58/4/044002}.
\newblock URL \url{https://doi.org/10.1088/0741-3335/58/4/044002}.

\bibitem[Ohshima et~al.(2021)Ohshima, Okada, Zang, Kobayashi, Minami, Kado, Adulsiriswad, Qiu, Matoike, Luo, Zhang, Miyashita, Motoshima, Nakamura, Konoshima, Mizuuchi, and Nagasaki]{ohshima2021ppcf}
S~Ohshima, H~Okada, L~Zang, S~Kobayashi, T~Minami, S~Kado, P~Adulsiriswad, D~Qiu, R~Matoike, M~Luo, P~Zhang, A~Miyashita, M~Motoshima, Y~Nakamura, S~Konoshima, T~Mizuuchi, and K~Nagasaki.
\newblock Isotope effect on zonal flow and its configuration dependence in low-density electron-cyclotron-resonance heated plasmas in heliotron j.
\newblock \emph{Plasma Physics and Controlled Fusion}, 63\penalty0 (10):\penalty0 104002, aug 2021.
\newblock \doi{10.1088/1361-6587/ac1837}.
\newblock URL \url{https://doi.org/10.1088/1361-6587/ac1837}.

\bibitem[Wakatani et~al.(2000)Wakatani, Nakamura, Kondo, Nakasuga, Besshou, Obiki, Sano, Hanatani, Mizuuchi, Okada, et~al.]{wakatani2000study}
M~Wakatani, Y~Nakamura, K~Kondo, M~Nakasuga, S~Besshou, T~Obiki, F~Sano, K~Hanatani, T~Mizuuchi, H~Okada, et~al.
\newblock Study of a helical axis heliotron.
\newblock \emph{Nuclear fusion}, 40\penalty0 (3Y):\penalty0 569, 2000.

\bibitem[Obiki et~al.(2001)Obiki, Mizuuchi, Nagasaki, Okada, Sano, Hanatani, Liu, Hamada, Manabe, Shidara, et~al.]{obiki2001first}
T~Obiki, T~Mizuuchi, K~Nagasaki, H~Okada, F~Sano, K~Hanatani, Y~Liu, T~Hamada, Y~Manabe, H~Shidara, et~al.
\newblock First plasmas in heliotron j.
\newblock \emph{Nuclear fusion}, 41\penalty0 (7):\penalty0 833, 2001.

\bibitem[Ohshima et~al.(2014)Ohshima, Kobayashi, and Yamamoto]{ohshima2014observation}
S~Ohshima, S~Kobayashi, and S~Yamamoto.
\newblock Observation of a toroidally symmetrical electric field fluctuation with radially elongated structure in heliotron j.
\newblock \emph{proceedings of 25th IAEA Fusion energy conference}, \penalty0 (EX/P4-26), 2014.

\bibitem[Ohshima et~al.(2006)Ohshima, Fujisawa, Shimizu, and Nakano]{ohshima2006observation}
Shinsuke Ohshima, Akihide Fujisawa, Akihiro Shimizu, and Haruhisa Nakano.
\newblock Observation of internal coherent mode structure using heavy ion beam probe.
\newblock \emph{Review of scientific instruments}, 77\penalty0 (10):\penalty0 10F517, 2006.

\bibitem[Crossley et~al.(1992)Crossley, Uddholm, Duncan, Khalid, and Rusbridge]{crossley1992experimental}
FJ~Crossley, P~Uddholm, P~Duncan, M~Khalid, and MG~Rusbridge.
\newblock Experimental study of drift-wave saturation in quadrupole geometry.
\newblock \emph{Plasma physics and controlled fusion}, 34\penalty0 (2):\penalty0 235, 1992.

\bibitem[Duncan and Rusbridge(1993)]{duncan1993amplitude}
PJ~Duncan and MG~Rusbridge.
\newblock The'amplitude correlation'method for the study of nonlinear interactions of plasma waves.
\newblock \emph{Plasma physics and controlled fusion}, 35\penalty0 (7):\penalty0 825, 1993.

\bibitem[Liu et~al.(2009)Liu, Lan, Yu, Zhao, Yan, Hong, Dong, Zhao, Qian, Cheng, et~al.]{liu2009characterizations}
AD~Liu, T~Lan, CX~Yu, HL~Zhao, LW~Yan, WY~Hong, JQ~Dong, KJ~Zhao, J~Qian, J~Cheng, et~al.
\newblock Characterizations of low-frequency zonal flow in the edge plasma of the hl-2a tokamak.
\newblock \emph{Physical review letters}, 103\penalty0 (9):\penalty0 095002, 2009.

\bibitem[Yamada et~al.(2008)Yamada, Itoh, Maruta, Kasuya, Nagashima, Shinohara, Terasaka, Yagi, Inagaki, Kawai, et~al.]{yamada2008anatomy}
Takuma Yamada, Sanae-I Itoh, Takashi Maruta, Naohiro Kasuya, Yoshihiko Nagashima, Shunjiro Shinohara, Kenichiro Terasaka, Masatoshi Yagi, Shigeru Inagaki, Yoshinobu Kawai, et~al.
\newblock Anatomy of plasma turbulence.
\newblock \emph{Nature physics}, 4\penalty0 (9):\penalty0 721--725, 2008.

\bibitem[Kim and Powers(1979)]{kim1979digital}
Young~C Kim and Edward~J Powers.
\newblock Digital bispectral analysis and its applications to nonlinear wave interactions.
\newblock \emph{IEEE transactions on plasma science}, 7\penalty0 (2):\penalty0 120--131, 1979.

\bibitem[Van~Milligen et~al.(1995)Van~Milligen, Sanchez, Estrada, Hidalgo, Bra{\~n}as, Carreras, and Garcia]{van1995wavelet}
B~Ph Van~Milligen, E~Sanchez, T~Estrada, C~Hidalgo, Bx~Bra{\~n}as, B~Carreras, and L~Garcia.
\newblock Wavelet bicoherence: a new turbulence analysis tool.
\newblock \emph{Physics of Plasmas}, 2\penalty0 (8):\penalty0 3017--3032, 1995.

\bibitem[Tynan et~al.(2001)Tynan, Moyer, Burin, and Holland]{tynan2001nonlinear}
GR~Tynan, RA~Moyer, MJ~Burin, and C~Holland.
\newblock On the nonlinear turbulent dynamics of shear-flow decorrelation and zonal flow generation.
\newblock \emph{Physics of Plasmas}, 8\penalty0 (6):\penalty0 2691--2699, 2001.

\bibitem[Nagashima et~al.(2005)Nagashima, Hoshino, Ejiri, Shinohara, Takase, Tsuzuki, Uehara, Kawashima, Ogawa, Ido, et~al.]{nagashima2005observation}
Y~Nagashima, K~Hoshino, A~Ejiri, K~Shinohara, Y~Takase, K~Tsuzuki, K~Uehara, H~Kawashima, H~Ogawa, T~Ido, et~al.
\newblock Observation of nonlinear coupling between small-poloidal wave-number potential fluctuations and turbulent potential fluctuations in ohmically heated plasmas in the jft-2m tokamak.
\newblock \emph{Physical review letters}, 95\penalty0 (9):\penalty0 095002, 2005.

\end{thebibliography}



\end{document}